\documentclass[aps,preprint,superscriptaddress]{revtex4}

\usepackage{graphics}
\usepackage{graphicx}
\usepackage{amssymb}
\usepackage{latexsym}
\usepackage{amsmath}
\usepackage{amsfonts}
\usepackage{color}

\newcommand{\atom}{_{\rm atom}}
\newcommand{\astr}{_{\rm astr}}

\begin{document}

\title{Proposal of an explanation of the Pioneer anomaly}
\author{Antonio F. Ra\~nada\footnote{Electronic address: afr@fis.ucm.es}}
\affiliation{ Facultad de F\'{\i}sica, Universidad Complutense,
28040 Madrid, Spain}
\author{Alfredo Tiemblo\footnote{Electronic address: Tiemblo@imaff.cfmac.csic.es}}
\affiliation{Instituto de Matem\'aticas y F\'{\i}sica Fundamental,
Consejo Superior de Investigaciones Cient\'ificas, Serrano 113b,
28006 Madrid, Spain}
\date{5 February 2009}

\begin{abstract}
We propose here an explanation of the Pioneer anomaly that is not in
conflict with the cartography of the solar system. In our model, the
spaceship does not suffer any extra acceleration but follows the
trajectory predicted by standard gravitational theory. The observed
acceleration is not real but apparent and has the same observational
footprint as a deceleration of astronomical time with respect to
atomic time. The details can be summarized as follows: i) as we
argued in a recent paper,  there is an unavoidable coupling between
the background gravitation that pervades the universe and the
quantum vacuum because of the long range and universality of
gravity; ii) via the fourth Heisenberg relation, we show that this
coupling causes a progressive desynchronization of the astronomical
and atomic clock-times, in such a way that the former decelerates
adiabatically with respect to the latter; and iii) since
gravitational theory uses astronomical time and the observers use
atomic time (they are using devices based on quantum physics), this
desynchronization
 necessarily  causes a  discrepancy between
 theory and observation, so that the observed velocity of the
 spaceship
  is smaller than the predicted one,
 in such a way that the Pioneer
 seems to lag behind its expected position. The discoverers of the anomaly suggested  ``the possibility
  that the origin of the anomalous signal is new physics" although
  they added
  ``the most likely cause of the effect
  is an unknown systematics," but also that ``In the unlikely event that there is new physics, one
  does not want to miss it because one had the wrong mind set."


\end{abstract}

\maketitle

\tableofcontents

\section{The Pioneer anomaly}
\textcolor{red}{This paper proposes a solution to the Pioneer
anomaly that is not in conflict with the cartography of the solar
system, which is up to now the most serious obstacle to the
explanation of this riddle. It is based on a previous paper by us,
in which we show that the coupling of the quantum vacuum and the
background gravitation of the entire universe must cause a
desynchronization of the astronomical and the atomic clocks that has
the same observational fingerprint as the Pioneer anomaly
\cite{RT08,Ran03}.}

This intriguing phenomenon, discovered in 1980
 and reported in 1998 by
Anderson {\it et al.} \cite{And98,And02}, consists of an
 adiabatic frequency blue drift of the two-way radio signals
from the Pioneer 10 and 11 (launched in 1972 and 1973), manifest in
a residual Doppler shift that increases linearly with time as
\begin{equation} \dot{\nu}/\nu =2a_{\rm t}\,,\quad \mbox{or}\quad
\nu =\nu_0\,[1+2a_{\rm t}(t-t_0)]\,,\label{10}\end{equation} where
$t_0$ is the launch time and $2a_{\rm t}= (5.82\pm 0.88)\times
10^{-18}\mbox{ s}^{-1}\simeq(2.53\pm 0.38) H_0$, $H_0$ being the
Hubble constant (overdot means time derivative). Many years after
its discovery, the phenomenon is still unexplained. As the
discoverers themselves explain  ``At that time [1980] the navigation
data had already indicated the presence of an anomaly in the Doppler
data; but at first the anomaly was only considered to be an
interesting navigational curiosity and was not seriously analyzed
\cite{Mar04}".

It must be stressed that the signal found by Anderson {\it et al.}
is very well-defined. Because of its linearity, the residual
frequency $\nu$ of the ship appears as a straight line in a plot
($\nu,\,t$) \cite{And98}. This strongly suggests that the anomaly is
the evidence of a new effect, probably with a cosmological origin,
so that the curve $\nu (t)$ can be accurately approximated by its
tangent for a scale of just a few decades. This contrasts with some
attempts to explain the effect as a local solar system phenomenon
\cite{Lam05,Rat06}. In 1998, after many attempts to explain the
effect, the discoverers admitted ``the possibility that the origin
of the anomalous signal is new physics" \cite{And98}, a suggestion
reiterated four years later in reference \cite{And02b}. But in the
conclusions of reference \cite{And02} (and in several other texts
where they repeat the same idea) they say with a pinch of caution
``Currently, we find no mechanism or theory that explains the
anomalous acceleration ... {\it Until more is known, we must admit
that the most likely cause of the effect is an unknown
systematics}." Consequently, they performed new analysis of the data
\cite{Tur06}, but they end their conclusions with this statement:
``{\it In the unlikely event that there is new physics, one does not
want to miss it because one had the wrong mind set}" (see also
\cite{Nie07}).

\subsection{A first interpretation of the effect}
Since it was detected as a Doppler shift that does not correspond to
any known motion of the spaceship, the simplest interpretation is
that there is an anomalous constant acceleration towards the Sun.
However, this is not acceptable since it would conflict with the
well-known cartography of the solar system and with the equivalence
principle.

Indeed, any interpretation of the anomaly
 in terms of a real, if unmodelled,
acceleration, due to a real force acting on the spaceship, would
face serious problems. Just as a simple example, the assumption that
the acceleration culprit is a distribution of dark matter around the
Sun with density $\rho \propto A/r$ and  a particular value of $A$,
gives the right value of the Pioneer acceleration for all the bodies
in the solar system. However, this would imply too much mass and
would predict an unacceptable variation in the perihelion of the
planets, increasing with the distance to the Sun. In fact, to
explain the anomaly without affecting the very well known
cartography of the solar system is probably the most serious
difficulty to solve the riddle. Anderson {\it et al.} included the
following paragraph in references \cite{And98} p. 2860 and
\cite{And02} p 41:

``The anomalous acceleration is too large to have gone undetected in
planetary orbits, particularly for Earth and Mars. NASA's  Viking
Mission provided radio-ranging measurements to an accuracy of 12 m.
If a planet experiences a small anomalous radial acceleration
$a_{\sc A}$, its orbital radius is perturbed by
       \begin{equation} \Delta r = -{\ell ^{\,6}a_{\sc A}\over (GM_\odot)^4}
\rightarrow - {ra_{\sc A}\over a_{\sc N}},\label{210}\end{equation}
where $\ell$ is the orbital angular momentum per unit mass and
$a_{\sc N}$ is the Newtonian acceleration at $r$. [The right value
in eq (\ref{210}) holds in the circular orbit limit.]

For Earth and Mars, this is about $-21$ km and $-76$ km. However,
the Viking data determine the difference between Earth and Mars
orbital radii to about 100 m accuracy and the sum to an accuracy of
150 m. The Pioneer effect is not seen [as it should be]."

In other words, the cartography of the solar system would be
affected in an unacceptable
   way.  However, it must be underscored that this irreproachable
argument by Anderson {\it et al.} assumes as a necessary condition
that the Pioneer effect is an anomalous but real
   acceleration, due to a real force therefore (the previous value of $\Delta r$  is calculated
   by adding $-a_{\rm A}$ to the RHS of the planets' radial equations of motion). If this is not so,
  the conclusion is not necessarily right and the effect does not
    necessarily include a negative correction to the radii of the
   orbits.

   This has motivated many to look for an unknown systematic
   error affecting the observations and has contributed to extending
   a misunderstanding: that it is not possible to solve the riddle without
 affecting the well-known solar system
   data. This is also the reason for the change in the name of the effect
   from ``Pioneer acceleration" to ``Pioneer anomaly."

   We interpret all this as a compelling indication
   that the solution will be found with a model in which the Pioneer
   acceleration is not real but only apparent. This is the case for
   our proposal.
   As will be shown in the following,  the
Pioneer follows, according to this work, the trajectory predicted by
the standard gravitation theory without any extra acceleration. Or,
more precisely, {\it the anomaly is not a real effect in our model,
but just an observational effect caused by the discrepancy between
the astronomical and atomic clock-times}.  It is, therefore, free
from this contradiction.

\subsection{A second interpretation of the effect}
Because of the previous problems, Anderson {\it et al.} then
attempted a second interpretation: $a_{\rm t}$ would be ``a clock
acceleration", expressing a kind of non-homogeneity of time
\cite{And98}. They imagined it in an intuitive and phenomenological
way, without theoretical reasons, explaining that, in order to fit
the trajectory, ``we were motivated to try to think of any {\it
(purely phenomenological) `time' distortions that might
fortuitously} fit the Pioneer results" (our emphasis, ref.
\cite{And02}, section XI-D). In particular, they added various
quadratic terms to the time, obtaining thus better fits to the
trajectory. In one,  which they call ``Quadratic time augmentation
model" they add to the TAI-ET (International Atomic Time-Ephemeris
Time) transformation the following distortion of the ephemeris time
ET
\begin{equation} \mbox{ET} \rightarrow \mbox{ET}^\prime=\mbox{ET} +{1\over
2}\,a_{\rm ET}\,\mbox{ET}^2\,,\label{20}\end{equation} where $a_{\rm
ET}$ is an inverse time. In the ``Quadratic in time model" which
they qualify as ``especially fascinating", they add a quadratic in
time term to the light time as seen by a DSN station, so that
\begin{eqnarray} \Delta \mbox{TAI}&=& \mbox{TAI}_{\rm
received}-\mbox{TAI}_{\rm sent} \label{22}\\
&\rightarrow& \Delta \mbox{TAI}'=\Delta \mbox{TAI}+{1\over
2}\,a_{\rm quad} (\mbox{TAI}_{\rm received}^2-\mbox{TAI}_{\rm
sent}^2).\nonumber
\end{eqnarray}
Note that to add quadratic terms is similar to introducing
desynchronizations or accelerations so that the times they use are
replaced by other differently defined times that accelerate with
respect to them. But they later gave up this idea because of the
lack of any theoretical basis and due to contradictions with the
determination process of the orbits.

\subsection{Our model} We argue here that the Pioneer anomaly is not a problem of
systematics or of data handling but a genuine new effect, an
indication in fact that we lack some important theoretical concepts.
More precisely, we show here that the anomaly has the same
observational signature as the desynchronization of the atomic and
the astronomical clocks caused by the coupling between the
background gravitational potential and the quantum vacuum, a new
phenomenon studied in \cite{RT08}, where the reader is referred to
for details (see also \cite{Ran03}).

In order to understand the meaning of the preceding sentence, let us
consider, as a simple illustrative example, the case in which an
observer has two different clocks that measure two times $t_1$ and
$t_2$, which accelerate adiabatically with respect to one another so
that ${\rm d} t_2/{\rm d} t_1=1+A(t-t_0)$, where $A={\rm
d}^2t_2/{\rm d} t_1^2$ is a small inverse time and $t_0$ an initial
time. Note that it is not necessary to specify the kind of time in
the RHS at first order. The clocks are non-equivalent because the
second derivative $A$ is non-zero. The observer measures the
velocity of the particle  with the two times, placing a series of
synchronized clocks of each kind along the particle trajectory and
observing the mark on the clocks when the position of the particle
coincide with that of each particular clock. He will thus find the
relation $v_2=v_1[1+A(t-t_0)]$, where the subindexes indicate the
time used to measure $v$. He would find the same effect on the
frequency of any spectroscopic line or even for the light speed, so
that $c_2=c_1[1+A(t-t_0)]$ and $\nu_2=\nu_1[1+A(t-t_0)]$ (compare
with (\ref{10})). If he were unaware of the  non equivalence of
$t_1$ and $t_2$, he would be surprised and would look for some error
in his measurement method (as a note of historical interest, we may
mention Milne's ideas which, on cosmological grounds, suggest the
possible existence of different time scales \cite{Mil37,Mil40}).

In spite of this simple example, all this may sound strange, but
note that it is improbable that the anomaly will involve only known
physics, considering that it remains unexplained more than 25 years
after its discovery. In other words, it would seem strange that,
when the solution of the riddle will be found, it would not seem
strange. To end this introductory section, it must be underscored
that, although the present model is based on new physical ideas, it
does not conflict with any established law or principle.

\section{The coupling background gravity -- quantum
vacuum and the desynchronization of the astronomical and the atomic
clocks} In a previous recent paper \cite{RT08}, we analyzed the idea
of a coupling between the background potential that pervades the
universe and the quantum vacuum, an obligatory effect due to the
universality and long range of gravitation. Building on the fourth
Heisenberg relation, we found that such a coupling would cause a
discrepancy between the two main clock-times of physics. The {\it
astronomical clock-time}, say $t_{\rm astr}$, is defined by the
trajectories of the planets and other celestial bodies, while the
{\it atomic clock-time}, say $t_{\rm atomic}$, is founded on the
oscillations of atomic systems. The former is measured with {\it
classical and gravitational} clocks, the solar system for instance,
while the latter is determined using {\it quantum and
electromagnetic} systems as clocks, in particular the oscillations
of atoms or masers. Contrary to the implicit tradition, these two
times could be different in principle since they are based on
different physical laws. They could be ticking at different rates,
even when located at the same place and having the same velocity.
They are certainly almost equal, at least at small scales, but since
we lack a unified theory of gravitation and quantum physics, the
assumption that they are exactly the same $t_{\rm astr} =t_{\rm
atomic}$ must not be accepted {\it a priori} without discussion. In
this work, the {\it march} of the latter with respect to the former
is defined as the derivative $u={\rm d}t_{\rm atomic} /{\rm d}t_{\rm
astr}$.

 To make this paper as self-contained as possible, here we will briefly summarize
 the main results of our previous paper \cite{RT08} (see also \cite{Ran03}.
 It is evident that the physics of the quantum vacuum is important because it fixes the
values of some observable quantities and gives rise to observable
phenomena. Probably its importance will increase during next
decades. We are interested in the sea of virtual electron-positron
pairs that pop-up and disappear constantly in empty space with their
charges and spins. On the average, speaking phenomenologically, a
virtual pair created with energy $E$
 lives during a time $\tau _0= \hbar /E$, according to the
fourth Heisenberg relation. This has an important consequence: the
optical density of empty space must depend on the gravitational
field. Indeed, if $\Psi $ is the averaged dimensionless
gravitational potential, {\it i. e.} the potential of all the
mass-energy in the entire universe assuming that it is uniformly
distributed, the pairs have an extra
 energy $E\Psi$, so that their lifetime and number density
 must depend on $\Psi$ as
 \begin{equation}  \tau _\Psi =\hbar/(E+E\Psi)= \tau
 _0/(1+\Psi);\qquad{\cal
N}_\Psi ={\cal N}_0/ (1+\Psi), \label{30}\end{equation} \cite{RT08}.
If $\Psi$ decreases, the quantum vacuum becomes denser, since the
density of charges and spins becomes higher; if $\Psi$ increases,
the quantum vacuum becomes thinner. In an expanding universe, the
background potential is time dependant $\Psi =\Psi (t)$ and
increases secularly and adiabatically so that its time derivative is
positive at present time $\dot{\Psi}_0
>0$ and small, probably of order $H_0$.
It must be stressed that {\it this  is not an {\em ad hoc}
hypothesis, but a necessary consequence of the fourth Heisenberg
relation and of the universality of gravitation}.

Let us take the Newtonian approximation and accept then the
following phenomenological hypothesis: the quantum vacuum can be
considered as a substratum, a transparent optical medium
characterized by a relative permittivity $\varepsilon _{\rm
r}(\Psi)$ and a relative permeability $\mu _{\rm r}(\Psi)$, which
are decreasing functions of $\Psi (t)$. As $\Psi(t)$ increases, the
optical density of the medium decreases (since there are less
charges and spins) and {\it vice versa}. As presented in previous
work, we express, at first order in the variation of $\Psi (t)$ and
near present time $t_0$, the permittivity and the permeability of
empty space as $\varepsilon =\varepsilon _0\varepsilon _{\rm
r}(t)=\varepsilon _0\{1-\beta [\Psi (t)-\Psi (t_0) ]\}$, $\mu
=\mu_0\mu _{\rm r}(t)=\mu _0\{1-\gamma [\Psi (t)-\Psi (t_0)]\}$
 where $\beta$ and $\gamma$ are certain coefficients, necessarily
 positive since the quantum vacuum must be dielectric but paramagnetic.
 The results of this paper will depend only
on the semisum $\eta =(\beta +\gamma)/2$.  It will be seen in the
following that the time $t$ is an astronomical time, as  is the
ephemeris time, so that
 we will note it as $t_{\rm astr}\,(=t)$ from now on.

 It was shown in former work \cite{Ran03} that, if the permittivity
 and the permeability vary adiabatically with time,
 the frequency and the speed of an
electromagnetic wave obey $ \dot{\nu}/
\nu=\dot{c}/c=-(\dot{\varepsilon}/\varepsilon +\dot{\mu}/\mu)/2=
\eta\dot{\Psi}_0$ (overdot means derivative with respect $t_{\rm
astr}$), if they are defined with respect to $t_{\rm astr}$, so that
near present time (or launch time) $t_{\rm astr,\,0}$
\begin{equation}  \nu_{\rm astr} =\nu _{\rm astr,\,0}[1+a(t_{\rm astr}-t_{\rm
astr,\,0})]\,;\qquad c_{\rm astr} =c[1+a(t_{\rm astr}-t_{\rm
astr,\,0})]\,, \label{40}\end{equation} with $a= \eta\dot{\Psi}_0\,
(>0)$.  If $a$ is very small, these variations are adiabatic. The
sub-index ``astr" indicates that the corresponding quantity has been
defined or measured using astronomical time $t_{\rm astr}$ and where
$c$ is the standard value of the light speed, the quantity that
appears in the tables of physical constants. We see that both the
frequency and the light speed increase adiabatically with time
$t_{\rm astr}$. In order to describe this situation, we define a
refractive index $n(t)=1+a(t_{\rm astr}-t_{\rm astr},\,0)$. However,
{\it $\nu$ and $c$ are constant if defined with the time
 $t _{\rm atomic}$}, determined by the relation
\begin{equation} {\rm d} t _{\rm atomic}=[1+a(t_{\rm astr} -t_{\rm astr,\,0})]\,{\rm d}
t_{\rm astr} =n(t_{\rm astr}){\rm d} t_{\rm astr}\,. \label{50}
\end{equation} It is clear that $t_{\rm atomic}$ is the time
measured by the atomic clocks, since the periods of the atomic
oscillations are obviously constant with respect to them. In fact
they are their basic units. The light speed is thus a universal
constant, as it must be, if defined with respect to $t_{\rm
atomic}$. Note that the derivatives with respect to the two
clock-times are equal at present time $t_{\rm astr,\,0}$ because
${\rm d}t_{\rm atomic}/{\rm d}t_{\rm astr}=1$ at that instant.
Hence, we can keep the same symbol for these two derivatives ${\rm
d} \Psi/{\rm d} t_{\rm astr}={\rm d} \Psi /{\rm d} t_{\rm
atomic}=\dot{\Psi}_0$. Since $\dot{\Psi}_0>0$ then $a>0$, so that
the march of the atomic clocks with respect to the astronomical ones
is equal to $u={\rm d}_{\rm atomic}/{\rm d}t_{\astr}=1+a(t_{\rm
astr} -t_{\rm astr,\,0})>1$ for $t_{\rm astr}>t_{\rm astr,\,0}$.
Note also that the refractive index $n$ is equal to the march $u$.

All this shows that the effect of the quantum vacuum would be to
desynchronize the astronomical and the atomic clocks in such a way
that the former decelerates adiabatically with respect to the
latter. As a consequence, the light speed and the frequencies would
increase progressively if they are measured with the astronomical
time $t_{\rm astr}$. However they are constant if measured with the
atomic time $t _{\rm atomic}$. Synchronizing the two times and
taking the international second as their common basic unit at
present time (or at launch time),  so that $t_{\rm astr,\,0}=t _{\rm
atomic,\,0}=t_0=0$, then ${\rm d} t_{\rm astr}={\rm d} t_{\rm
atomic} $ (ref. \cite{RT08}).

 It follows from
(\ref{50}) that, if $t_0=0$,
\begin{equation}  t_{\rm astr}=t_{\rm atomic} -{1\over 2}\, a\,t_{\rm
atomic} ^2 \qquad \mbox{ and }\qquad {\!\!{\rm d} ^2t_{\rm
astr}\over {\rm d} t_{\rm atomic} ^2}=-\eta \dot{\Psi}_0
=-a\label{60}\,.\end{equation} There is a striking similarity
between (\ref{60}) and (\ref{20})-(\ref{22}), {\it which could
explain why Anderson {\it et al.} obtained improved data fits by
introducing these time distortions}. As can be seen, the quantity
$a=\eta\dot{\Psi}_0$ in our model corresponds to $2a_{\rm t}$ in the
papers by Anderson {\it et al.}, which shows that it is $a$ or
$2a_{\rm t}$, not $a_{\rm t}$, which truly deserves the name of
``clock acceleration".

If the march $u={\rm d}t_{\rm atomic}/{\rm d}t_{\rm astr}=1+a(t_{\rm
astr}-t_{\rm astr,\,0})$ is not constant, the two clock-times are
not equivalent, so that the speed measured using the Doppler effect
with devices sensitive to the quantum time, say $v_{\rm atomic}={\rm
d} \ell /{\rm d} t_{\rm atomic}$, would be different from the
astronomical speed $v_{\rm astr}={\rm d} \ell /{\rm d} t_{\rm
astr}$. Indeed
\begin{equation} v_{\rm atomic}= v_{\rm astr}/u=v_{\rm astr}[1-a(t_{\rm astr}-t_{\rm
astr,\,0})]\,,\label{80}\end{equation} and the frequencies obey the
same relation $\nu_{\rm atomic}= \nu_{\rm astr}[1-a(t_{\rm
astr}-t_{\rm astr,\,0})] $. Since gravitational theory gives $v_{\rm
astr}$ while the observers measure $v_{\rm atomic}$, {\it there must
be a discrepancy between theory and observation}. In this way, an
apparent but unreal violation of standard gravity would be detected.

\section{Explanation of the anomaly} These arguments give a
compelling explanation of the anomaly as an effect of the
discrepancy between these two times. Let us see why. The frequencies
measured by Anderson {\it et al.} are standard frequencies defined
with respect to atomic time $t_{\rm atomic}$, since they observe
them with devices based on quantum physics. {\it They did not
measure frequencies with respect to the astronomical time}. However,
since (i) the trajectory was determined by standard gravitational
theory that use astronomical time and (ii) the observed frequency
shifts used atomic time, they found a discrepancy between theory and
observation: this is the Pioneer Anomaly.

Once all this is  accepted, the anomaly is easily explained. To do
that, we will now follow two paths, both very simple: first a more
intuitive and geometrical explanation, and second a more precise and
formal account. The first starts with the distance traveled by a
spaceship along a given trajectory, which can be expressed in two
forms
\begin{equation} d=\int _0^{t_{\rm atomic}}\!\!\!\!\!\!\!\!v_{\rm atomic}(t')
{\rm d}t'=\int _0^{t_{\rm astr}}\!\!\!\!\!v_{\rm astr}(t')\,{\rm d}
t'\,,\label{110}\end{equation} where $t'=t_{\rm atomic}$ in the
first integral, $t'=t_{\rm astr}$ in the second and $(0,t_{\rm
atomic})$ and $(0,t_{\rm astr})$ are the same time interval
expressed with the two times (because $v_{\rm atomic} =v_{\rm
astr}/u$ and ${\rm d}t_{\rm atomic}=u{\rm d}t_{\rm astr}$). Equation
(\ref{110}) is always valid, whether or not the two times are equal.
 What we are proposing as the solution to the riddle is that the
two times are in fact different. If, however, they are assumed to be
equal $t_{\rm astr}=t_{\rm atomic}$, then the distance deduced from
observations $d_{\rm observ}$ and the expected distance according to
standard gravitational theory $d_{\rm theory}$ up to the same
$t_{\rm atomic}$, which are
\begin{eqnarray} d_{\rm observ}&=&\int _0^{t_{\rm atomic}}\!\!\!\!\!\!\!\!\!\!\!\!v_{\rm
atomic}(t')\,{\rm d}t'\,,\quad d_{\rm theory}=\int _0^{t_{\rm
atomic}}\!\!\!\!\!\!\!\!\!v_{\rm astr}(t')\,{\rm d} t'\,,\nonumber
\\ & &\label{120}\end{eqnarray} with $t'=t_{\rm atomic}$ in both integrals, would be different.
Now, if $u>1$ as it was during the Pioneer flight, the atomic
velocity is smaller so that $d_{\rm observ}<d_{\rm theory}$: {\it it
would seem that the ship travels a smaller than expected distance}.
Apparently, it would lag behind the predicted position.

\begin{figure}[h]
\begin{center}
\scalebox{1}{\includegraphics{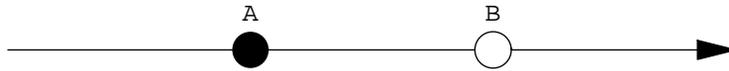}}
\end{center}
\caption{Schematic representation with arbitrary units of the
 Pioneer anomaly (explanation in the text.)} \label{Fig1}
\end{figure}

All this is explained in Figure \ref{Fig1}, where  the Pioneer
trajectory receding from the Sun is plotted schematically. The
spacecraft moves in the direction of the arrow. The white circle at
$B$ is the position of the ship according to standard gravitation,
its real position in fact if this theory is correct. The black
circle at $A$ is the apparent position, deduced from the ship's
velocity after measuring the Doppler effect on the frequencies of
the signals using atomic clocks and devices. If the two times were
the same, as is usually assumed, $A$ and $B$ would coincide; if they
are different, as in this model, $A$ and $B$ separate, $A$ being
only an apparent position.  Which one of the pair $A,\,B$ moves
faster depends, therefore, on the value of the relative march of the
atomic clocks with respect to the astronomical clocks $u={\rm d}
t_{\rm atomic}/{\rm d}t_{\rm astr}$.

In the case of the Pioneer, $u>1$ so that $v_{\rm atomic}<v_{\rm
astr}$: the apparent position $A$ lags increasingly behind the real
position $B$. The consequence is an unexpected blue Doppler
residual, easily interpreted as an extra anomalous acceleration
towards the Sun, the so-called Pioneer acceleration. However, the
Pioneer suffers no real acceleration due to any kind of
gravitational force. The anomaly is simply the manifestation of the
desynchronization of the two kinds of clocks.

It is interesting to examine this argument following a second path
(for simplicity, we will omit in this next argument the subscripts
``atom" and ``astr" in $\nu$, $v$ or $t$, whenever it is not
necessary to distinguish between the two cases at first order). Let
a team of physicists perform Doppler observations of the two-way
signals to and from a spaceship. If their devices are not of high
precision, their expected and observed values will both be equal to
$\nu ^{\rm \,obs}=\nu ^{\rm \,em}(1-2v/c)$, in accordance with
elementary standard textbooks, where $v$ is the recession velocity
of the ship. Let us now assume that their instruments are very
precise and that the ship trajectory can be determined with very
high accuracy, as is the case with the Pioneers 10 and 11. If the
two times are different but the observers are unaware of it, they
will expect to find the value $\nu_{\rm expected}=\nu _{\rm
atom}^{\rm \,em}(1-2v/c)$, so written because they use atomic time.
However, what they will observe in fact with their measuring devices
based in quantum physics is the value predicted by gravitation
theories $\nu _{\rm observed}=\nu _{\rm astr}^{\rm \,em}(1-2v/c)$,
expressed in terms of atomic clock-time, {\it i. e.}
\begin{equation} \nu _{\rm observed}=\nu _{\rm atom}^{\rm em}[1+a(t-t_0)]\,(1-2v/c)\,=\nu
_{\rm atom}^{\rm em}[1-2v/c+a(t-t_0)].\label{90} \end{equation}
This means that, in addition to the Doppler frequency shift
$-2v/c$, they will find an unexpected Doppler residual towards the
blue, increasing in time as  $a(t-t_0)$. In self-explaining
notation, they would write
\begin{equation}{\nu _{\rm observed}-\nu _{\rm expected}\over \nu _{\rm
expected}}=a(t-t_0),\label{100}\end{equation} at first order, which
is the same as (\ref{10}) if $a=2a_t$ (if the RHS of (\ref{100}) is
interpreted  as a standard Doppler effect, it would be written as
$2a(t-t_0)$). Although the observations were received with surprise,
they are what should have been expected, had the discrepancy between
the two times been an accepted effect in 1980. This argument shows
that the effect of the desynchronization of the times and the
Pioneer anomaly coincide qualitatively. They are probably the same
phenomenon.

 The preceding arguments explain the statement made in  section 1 that the Pioneer anomaly has the same
observational footprint as a desynchronization of the astronomical
and the atomic clocks such that the former decelerate with respect
to the latter. Note that the failure to include the non-zero
acceleration ${\rm d} ^2t_{\rm atomic}/{\rm d} t_{\rm astr}^2$ in
the analysis mimics a blue Doppler residual $\dot{\nu}/ \nu= a$ or
$\nu =\nu _0[1+a(t_{\rm astr}-t_{\rm astr,\,0})]$. This is precisely
what Anderson {\it et al.} observed and gives the right result if
$a\simeq 2a_{\rm t}$, so that $\eta \dot{\Psi}_0\simeq 2.5 H_0$.

For this model to be right, it is necessary that $\dot{\Psi} _0>0$,
i.e. that the present value of the time derivative of the background
potential be positive. Indeed, it could be argued that this can be
considered in fact a prediction of the model. Alternatively, a
simple argument shows that $\Psi(t)$ must be increasing at present
time. In fact, it is the sum of two terms, one due to the matter,
either ordinary and dark, and another due to the cosmological
constant or the dark energy. The potential of a set of masses grows
if they separate. In the case of the visible universe, however, the
total mass is increasing since new galaxies enter constantly through
the particle horizon, continuously adding negative potential. For
this reason the first term is not necessarily increasing. The second
term, on the other hand, is always positive, grows with the
expansion and becomes dominant since  a look-back time near at least
$t_0/2$, when the set of galaxies passed from the slow-down to the
speed-up. More precisely, the potential of the two terms are
approximately proportional to $-1/S$ and $+S^2$ at present time,
respectively, where $S$ is the scale factor.

\section{The Pioneer anomaly and the planets}

The prediction that the anomaly is not due to any acceleration is an
important and promising feature of our model, since it is free,
therefore, of any criticism about its effect on the planetary
orbits. What we are proposing is that the Pioneers 10/11 did not
suffer any extra acceleration but, quite on the contrary, that they
followed the standard gravitation theory, and that the observed and
unmodelled acceleration was an observational effect of the
desynchronization of the atomic and the astronomical clocks due to
an interaction between the background gravity and the quantum
vacuum. This means that our model is not in conflict with the
cartography of the Solar system.

In spite of the previous comments, it must be said that in our model
the anomaly does affect the radiation from the planets; in fact, it
must affect any signal sent to or coming from any celestial body.
This does not mean that their orbits change, just that there is a
difference between the expected and the observed frequencies coming
from them. But it will be extremely difficult to observe these
variations in the case of the planets because the effect on them is
poorly defined, blurred and too small to be detected at the present
time. To have been able to observe and measure the Pioneer anomaly
was an extraordinary achievement by Anderson {\it et al.}. Their
discovery was nonetheless facilitated by the simplicity of a ship
like the Pioneer as compared with the planets. Any change in the
observational conditions or in the craft  could well have made such
measurements impossible or, at least, less accurate for the planets.
The following ideas that come quickly to mind will serve to
illustrate the extreme difficulty of making similar observations on
the
planets.\\
$\bullet$  It would be very difficult, to say the least, to send a
microwave signal from earth and then  detect the reflected
one. There are no transponders on the planets.\\
$\bullet$ As with the Pioneer, the solar wind would perturb the
signal to the point of making it useless until at least the orbit of
Saturn at 9.5 AU. The NASA team started to obtain significant
results for the Pioneer at 20 AU, when the probe was crossing the
orbit of Uranus .\\$\bullet$ The ionospheres of the outer planets
would interact strongly with the signal, whereas the Pioneer has no
ionosphere.\\ $\bullet$  Unlike the Pioneer, the outer planets emit
stochastic radiofrequencies and microwaves that fluctuate and blur
the signal. \\
$\bullet$ The outer planets are gaseous thus lacking a well defined
surface while the Pioneer has a perfectly defined border. This would
blur the reflected signal.
\\$ \bullet$  In the light rays emitted from any planet, there is surely a
residual frequency toward the blue verifying, as in the case of the
Pioneer, the relation  $\nu\astr =   \nu\atom   [1+a(t - t_0)]$,
where $\nu\atom = \nu_{\rm obs}$ and   $\nu\astr =  \nu_{\rm exp}$
are, respectively, the atomic (observed) and astronomical (expected)
frequencies and $a = 2 a_{\rm t}$, is twice what Anderson {\it et
al.} called ``a clock acceleration". If this relation were due to a
Doppler effect, it could be written as  $\nu\atom = \nu\astr (1- 2
v_{\rm P}/c)$, where $v_{\rm P}$ is the ship velocity. Taking the
temporal derivative, it follows that $v_{\rm P}= v_{\rm P,0} - ac (t
- t_0)/2$, so that the Pioneer would have an unmodelled
 acceleration $a_{\rm P} = ac/2$ toward the Sun. Anderson {\it et al.} determined its value
 within an error of 15 $\%$ (i. e. 1.33/8.74), which means that this is also the error in
 $a$.

    But there are several additional and relevant differences between
how the anomaly might be observed for the planets and for the
Pioneer. The ship velocity is greater than about 11 km/s (its
approximated asymptotic speed) and is monotonously directed away
from the Sun; the effect on the Pioneer is therefore secular. On the
other hand, the radial velocities of the planets are periodic and
much smaller. Looking from the Sun and taking the approximation of
circular orbits, there would be no effect, since the radial
velocities verify $\dot{r}\atom = \dot{r}\astr   [1- a (t - t_0)]$
and the astronomical radial velocity $\dot{r}\astr$ vanishes, where
the overdot denotes time derivative. With elliptic orbits, the
effect on the radial velocity must vanish at the apsidal points and
be small for small eccentricities. The radial velocity is positive
from a perihelion to the following aphelion and negative from this
aphelion to the next perihelion. Its average value is zero.  The
maximum value of the astronomical radial velocity of the planet
$\dot{r}\astr$ follows from the elementary theory of planetary
motion,
 \begin{equation}  \dot{r} _{\rm astr,\,max,\,planet}= \left({GMe^2\over a(1-e^2)}\right)^{1/2},\label{300}\end{equation}
 $a$ and $e$ being the semi-major axis and the eccentricity (we use only this notation from here to eq. (\ref{320}),
 for the rest of the paper $a$ is the relative acceleration of the clock-times).

The extra radial velocity is thus $ \dot{r}\atom - \dot{r}\astr = -
\dot{r}\astr a (t - t_0)$. So,  the relative intensity of the effect
on the radial velocity of a planet with respect to the Pioneer
anomaly would be
\begin{equation} \Theta ={\dot{r}_{\rm astr,\, max,\,planet}\times a (t -
t_0) \over \dot{r}_{\rm astr, Pion}\times a (t - t_0)} =
{\dot{r}_{\rm astr,\,max,\, planet} \over \dot{r}_{\rm astr,\,
Pion}},\label{310}\end{equation} {\it  i. e.} the ratio of the
radial velocities. For the Pioneer we take now its asymptotic
velocity, which is smaller than the one during the observations.
With the previous formulae and the values of the semi-major axis and
eccentricities, it is easy to find the value of $\Theta$ for the
different planets. In self-explaining notation, they are,
approximately,
\begin{eqnarray} &&\Theta _{\rm Mer} = 0.9;\quad   \Theta _{\rm Ven} = 0.02; \quad \Theta
_{\rm Earth}=  0.04;\quad \Theta _{\rm Mar} = 0.2;\nonumber \\
&&\Theta _{\rm Jup}= 0.06;\quad \Theta _{\rm Sat} = 0.05;\quad
 \Theta _{\rm Ura} = 0.03;\quad  \Theta _{\rm Nep} = 0.005.\label{320} \end{eqnarray}

    In order to measure the effect, it would be necessary to measure
velocity differences much smaller than those of the Pioneer. The
fact that the error in the Pioneer anomaly is about 15 $\%$ means
that an effect smaller than 15 $\%$ would be too small to be
measured. Though Mercury has a ratio nicely above this value
(because it is near the Sun and its eccentricity is large) and the
Mars ratio is near the borderline of detection, both these planets
will have the effect blurred by the solar wind, particularly for
Mercury. Remember that the anomaly began to be seen clearly at 20
AU, about at the orbit of Uranus. The values of for this planet and
for Neptune are small, 0.03 and 0.005. We conclude, therefore, that
the effect on the planets, although existent in our model, is too
small and too ill defined to be detectable with current technology.

\bigskip

\section{Summary and conclusions}

It was shown in previous work \cite{RT08} that, because gravitation
is long range and universal, affecting all kinds of mass or energy,
a coupling necessarily  exists between the background gravitation
that pervades the universe and the quantum vacuum. This coupling can
be estimated using the fourth Heisenberg relation and implies a
progressive attenuation of the quantum vacuum in the expanding
universe, if measured with astronomical clocks. This means that the
density of virtual pairs and the optical density of empty space
decrease as well. In its turn, this causes a desynchronization of
the astronomical and atomic clocks which gives a solution of the
Pioneer anomaly as a quantum cosmological effect. Therefore

 (i) The
Pioneer anomaly (\ref{10}) can be understood as the adiabatic
decrease in the periods of the atomic oscillations, if measured with
astronomical time or, equivalently, as the deceleration of the
astronomical clock-time with respect to the atomic clock-time, equal
to twice what Anderson {\it et al.} called the ``clock
acceleration", $2a_{\rm t}={{\rm d} ^2t_{\rm atomic}/ {\rm d} t_{\rm
astr}^2}|\,_0>0$. This quantity can be expressed as $2a_{\rm t}=\eta
\dot{\Psi}_0$, where $\dot{\Psi}_0$ is the present time derivative
of the background potential and $\eta$ a coefficient that represents
the variation of the properties of the vacuum. In other words, the
atomic clocks speed up adiabatically with respect to the
astronomical clocks. This would be  a certain evidence
 of the interplay between gravitation
and quantum physics.  In other words, we are proposing here that
what Anderson {\it et al.} observed is the relative march of the
atomic clock-time of the detectors with respect to the
 astronomical clock-time of the orbit $u=1+\eta \dot{\Psi}_0(t_{\rm astr}-t_{\rm astr,\,0})$
 (compare with (\ref{10})). Alternatively, they observed the
 variation of the refractive index $n(t_{\rm astr})$ along the Pioneer trajectory.
What should be labeled ``clock acceleration" is not $a_{\rm t}$ but
rather $2a_{\rm t}$. Although this new idea may seem surprising and
strange, it conflicts with no physical law or principle.

(ii)  In order to know whether this explanation is quantitatively
right, it is necessary to estimate the value of the ``clock
acceleration" $2a_{\rm t}$. However, this value depends on a
coefficient, here called $\eta$, which cannot be calculated without
a theory of quantum gravity. On the other hand, the Pioneer anomaly
could be considered as a measurement of $2a_{\rm t}$ to be used in
the future as a test for quantum gravity. In any case, it was shown
in the previous paper \cite{RT08} that the deceleration of $t_{\rm
astr}$ with respect to $t_{\rm atomic}$ does not affect the
experimental values of the spectral frequencies, the periods of the
planets or the gravitational redshift, because these quantities are
all measured with devices that use atomic time.

Some final comments. First, as a consequence of the coupling between
the background gravitation and the quantum vacuum, the light speed
would increase with an acceleration $a_\ell =2a_{\rm t}c$, equal to
twice the so called Pioneer acceleration \cite{And98,And02}, if
defined or measured with respect to the astronomical clock-time.
However it is constant if measured using atomic clock-time. But note
that astronomical time is never used to measure the frequency of an
electromagnetic wave or the speed of light. In fact, an atomic clock
is the ``natural clock" to define and measure the light speed, since
its basic unit is the period of the corresponding electromagnetic
wave, so that the speed and the frequencies are then necessarily
constant. This means that $c$ is still the fundamental constant we
know if measured with atomic time.

 Second, since the Pioneer anomaly would be a quantum effect,
which causes the light speed and the frequency to increase if
defined and measured with astronomical proper time, it would be
alien to general relativity. It must also be stressed that, if we
accept that there are non-equivalent clocks that accelerate with
respect to one another because of a coupling between gravity and the
quantum vacuum, then a new field of unexplored physics opens. In
particular, the role of the parametric invariance in our description
of the cosmological problems.

This discrepancy between theory and observations could also affect
the Hubble law.  A galaxy has at least a couple of things in common
with the Pioneer: both are receding from us and both their
velocities are Doppler measured with quantum devices that use atomic
time to compare their values with the predictions of a gravitational
theory based on astronomical time, the Friedmann equation for the
galaxies and the theory of orbits for the Pioneer. This is a
question that merits in-depth analysis.

\section{Acknowledgements.} We are grateful to profs. A. I. G\'omez de Castro, J. Mart\'{\i}n
 and J. Us\'on for helpful discussions.

\end{document}